\documentclass[a4paper]{jpconf}
\usepackage{graphicx}
\begin{document}
\title{Tevatron Top-Quark  Combinations and World Top-Quark Mass Combination}

\author{Reinhild Yvonne Peters, on behalf of the ATLAS, CDF, CMS and
  D0 collaborations}

\address{University of Manchester, School of Physics and Astronomy,
  Oxford Road, Manchester M13 9PL, England; also at DESY, Hamburg, Germany}

\ead{reinhild.peters@cern.ch}
\begin{abstract}
Almost 20 years after its discovery, the top quark is still an
interesting particle, undergoing precise investigation of its properties. For many years, the
Tevatron proton antiproton collider at Fermilab was the only place to
study top quarks in detail, while with the recent start of the LHC proton proton
collider a top quark factory has opened. An important ingredient for
the full understanding of the top quark is the combination of
measurements from  the individual experiments. In particular, the Tevaton
combinations of single top-quark cross sections, the $t\bar{t}$ production
cross section, the $W$ helicity in top-quark decays as well as the
Tevatron and the world combination of the top-quark mass are
discussed.  
\end{abstract}

\section{Introduction}
Since its discovery in 1995 by the CDF and D0 experiments at the
Fermilab Tevatron proton antiproton
collider~\cite{cdftopdiscovery}\cite{d0topdiscovery}, the top quark
has been studied in detail by four different experiments: CDF and D0
at the Tevatron, and since 2009 also by the ATLAS~\cite{atlas} and
CMS~\cite{cms} experiments at the LHC proton proton collider at
CERN. While each collaboration performs their measurements independent
of the other experiments, an important step for the full understanding
of the top quark is the combination of the acquired measurements
between experiments at the same collider, as well as experiments at
the two different colliders for those properties that are independent of
collision energy or type. 

In the following  I present the combinations of top-quark  measurements between CDF
and D0 as well as the first world combination of the top-quark
mass. The Tevatron combinations are the recent combination of single
top-quark measurements, the top quark-antiquark ($t\bar{t}$) production
cross section, the $W$ helicity and the top-quark mass. 

\section{Tevatron Combinations}
Since the discovery of the top quark, intense programs to study the
heaviest known elementary particle have been undergone at both
experiments, CDF and D0, and are partially still going on today. The
Tevatron Runs in which top quarks were studied were Run~I and
Run~II. Run~I lasted form 1992 to 1996, providing about 120~pb$^{-1}$
of $p\bar{p}$ collisions at an energy of 1.8~TeV.
In 
Run~II $p\bar{p}$ collisions at an energy of 1.96~TeV took place  (starting 2001
and ending September 30th, 2011),
providing about $10.5$~fb$^{-1}$ of integrated luminosity for
each of
the D0 and CDF experiments. Several analyses used in the combination
are based on the full Run~II data sample.

\subsection{Single Top Cross Section Combinations}
The two top-quark production mechanisms are production in pairs via the strong interaction, or
singly via the electroweak interaction. The latter occurs via
$s$-channel, $t$-channel and $Wt$-channel production. While the
$Wt$~channel has a negligible production cross section at the
Tevatron, the other two channels have now all been observed at the
Tevatron. In particular, the observation of $s$-channel single top
production was only possible via the combination of the CDF and D0
analyses.
The measurement of  single top quark
production is very challenging, as the main background from $W$+jets
events looks very similar to the single top signature. Various
multivariate techniques have been employed to distinguish the signal
from the large background.
For the observation of $s$-channel single top, events with one
isolated, high-transverse-momentum ($p_T$) electron or muon, large
missing transverse energy from the not-detected neutrino and two (or
two and three for D0)
high-$p_T$ jets, of which one or two have to be identified as $b$-jets, are
considered in both collaborations. In addition, CDF also performs an
analysis, where
events with a missing transverse energy plus jet signature are used,
adding events to the sample in which the lepton is not directly reconstructed.
A multivariate
discriminant is built to separate $s$-channel signal from
background. The combined cross section is then extracted by performing
a Bayesian statistical analysis, where the likelihoods of the analyses
from both collaborations are combined. The systematic uncertainties are
categorised in classes, taking into account the correlations of
different sources between the analyses and between the
experiments. The central result is taken as the maximum of the
posterior probability.  In this analysis, the $t$-channel single top production
cross section was
set to its standard model (SM) value.  The combined analysis results in a cross section of
$\sigma_{s}=1.29^{+0.26}_{-0.24}$~pb~\cite{tevschannel}, which deviates with more than  6.3 standard
deviations  from zero and is consistent with the SM prediction.

\begin{figure*}[t]
\centering
\includegraphics[width=55mm]{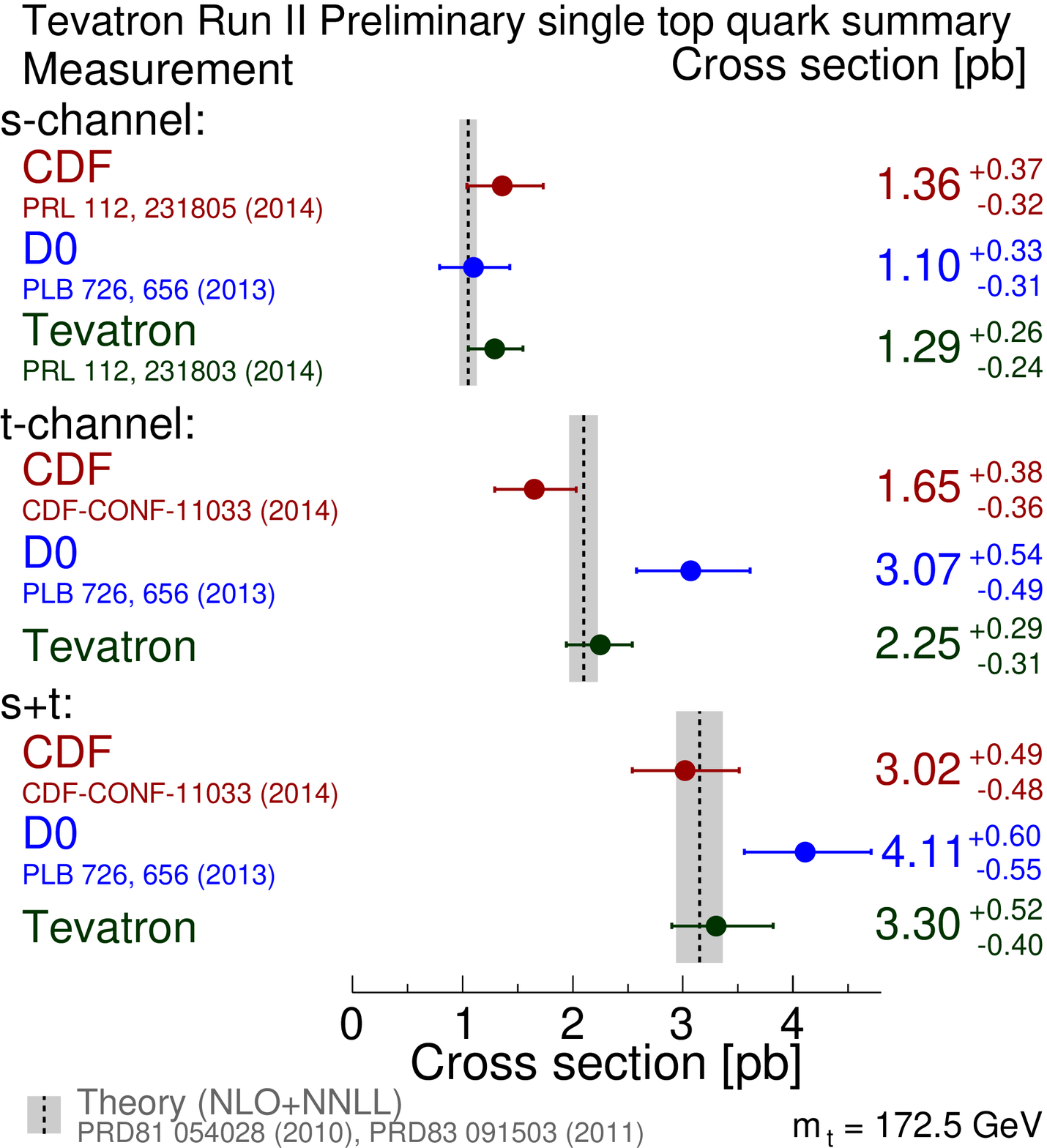}
\includegraphics[width=68mm]{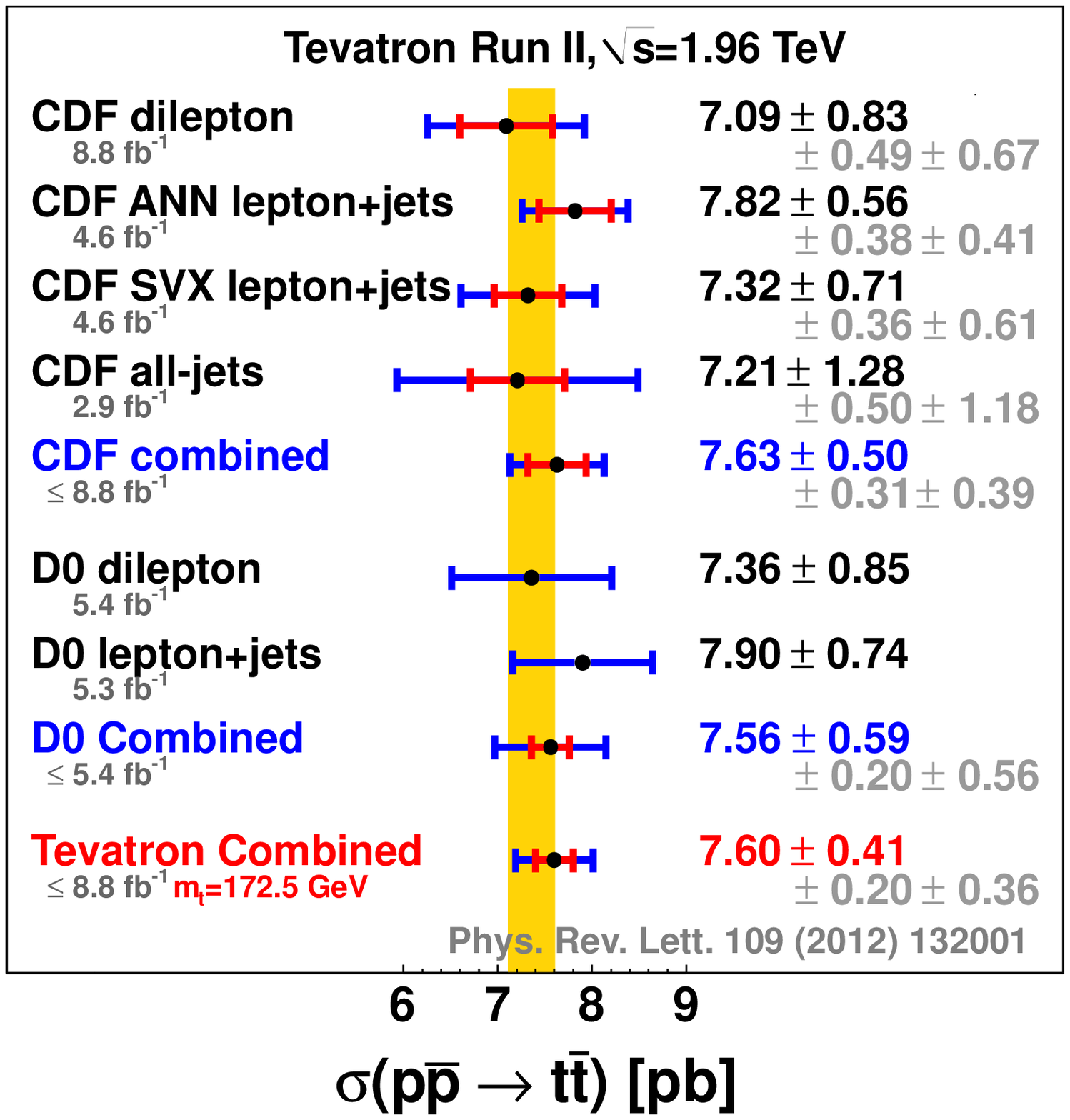}
\caption{ Recent Tevatron combination of single top quark production
  cross sections (left)~\cite{tevsinglecombis} and $t\bar{t}$
  production cross sections (right)~\cite{tevxseccombi}.} \label{fig:xseccombi}
\end{figure*}

The same methods and input analyses have also been used for a
combination of the $t$-channel and $s+t$-channel cross section. For
the  $s+t$-channel  combinations, the multivariate discriminants
were trained separately on
$s$-channel and $t$-channel as signal, and both discriminants were
used simultaneously. A 2D posterior probability density of
$sigma_{s}$ versus $sigma_{t}$ was then constructed for
the $s+t$-channel measurement. For the
extraction of $sigma_{s+t}$, the $t$-channel is then
integrated out. The resulting cross sections are
$\sigma_{t}= 2.25^{+0.29}_{-0.31}$~pb and
$\sigma_{s+t}=3.30^{+0.52}_{-0.40}$~pb~\cite{tevsinglecombis}.
Figure~\ref{fig:xseccombi} (left) shows an overview of all Tevatron
combinations of single top quark processes. 

Using the same discriminants, also the square of the CKM-matrix
element, $|V_{tb}|^2$, can be extracted. In this case the Bayesian
posterior probability density is formed for $|V_{tb}|^2$, using a flat
prior in $|V_{tb}|^2$. The result yields
$|V_{tb}|^2=1.04^{+0.12}_{-0.10}$~\cite{tevsinglecombis}.

\subsection{$t\bar{t}$ Cross Section Combinations}
At the Tevatron, the production of top-quark pairs is dominated by
$q\bar{q}$ annihilation with a fraction of approximately 85\%, while
gluon-gluon fusion contributes with about $15$\%. 
The Tevatron combination of the $t\bar{t}$ cross section uses as input
four analyses by CDF and two by D0. In particular, the CDF analyses
are: a counting-method in dileptonic final states, where events with
at least one identified $b$-jet are considered; an analysis in the
semileptonic final state, where a neural-network discriminant is built
based on various kinematic variables; another analysis in semileptonic
events, where events with at least one identified $b$-jet are counted;
and an analysis in all-hadronic events, where a fit to the
reconstructed top-quark mass is performed on events with six to eight
jets of which one or more than one are identified as $b$-jets. In a
first step, these CDF analyses are combined using the BLUE
method~\cite{blue}, taking into account
correlations between analyses and systematic uncertainties. Since the
systematic uncertainty on the selection efficiency is directly
proportional to the measured cross section, three iterations of BLUE
are performed in order to remove any potential bias. 

Furthermore, the two analyses by D0 are also first combined
experiment-internally, where the combination is done by a likelihood
fit, treating the systematic uncertainties as nuisance
parameters. The two analyses are one in the dilepton final state,
where a fit to the discriminant based on the neural network
$b$-jet identification algorithm is performed, and an analysis in the
semileptonic final state, in which events with three or more than
three jets, split into samples of zero, one or at least two identified
$b$-jets, are either counted or used to build a random forest
discriminant. The two D0 and CDF combinations are then further
combined using BLUE, with a more rough classification of the
systematic uncertainties. The final combined cross section is
$\sigma_{t\bar{t}}= 7.60 \pm 0.41$~pb~\cite{tevxseccombi}, which is consistent with the
standard model (SM)
prediction. The compatibility between the cross sections by CDF and D0
is 17\%. Figure~\ref{fig:xseccombi} (right) shows the $t\bar{t}$
cross section combination and all individual inputs.

\subsection{ $W$-Helicity Combination}
The very first Tevatron combination of a top-quark property has been
the combination of the $W$-helicity measurements in $t\bar{t}$
events, using up to 5.4~fb$^{-1}$. Since $W$~bosons couple purely left-handedly to fermions in the
SM, the relative orientation of the spin of the $b$-quark and the
$W$~boson from the top-quark decay are constrained. The fractions
of negative ($f_{-}$), zero ($f_0$) and positive ($f_{+}$) helicity of
the $W$ boson are predicted to be $f_{-}=0.685\pm0.005$,
$f_0=0.311\pm0.005$ and $f_{+}=0.0017\pm0.0001$ at
next-to-next-to-leading order (NNLO) QCD~\cite{nnloheli}.

The combination of the $W$-helicity includes two measurements from CDF
and one analysis from D0. In particular, one of the input analyses
performed by CDF uses the Matrix Element method in semileptonic
$t\bar{t}$ events. In this method, matrix elements for different
assumptions of $f_0$ and $f_{+}$ are calculated, integrated with
parton distribution functions and transfer functions, that map the
true momenta of the particles used in the matrix element to the
measured quantities, and are then compared to data. The second input analysis by CDF is
based on a template method, where templates in  the cosine of the angle $\theta^{*}$ between the
down-type decay product of the $W$ boson and the top quark in the $W$ boson
rest frame are formed for the different helicity states and fitted to data. The analysis is performed on semileptonic events. The D0
analysis uses the same strategy of a template fit, which is done in
dileptonic and semileptonic events.  All analyses are performed by fitting the fractions $f_0$ and $f_{+}$ simultaneously,
only constraining the sum of all three fractions to be one. Using this model independent approach, the combination yields
$f_0=0.732 \pm 0.063 {\rm (stat)} \pm 0.052 {\rm (syst)}$ and
$f_{+}=-0.039 \pm 0.034 {\rm (stat)} \pm 0.030 {\rm (syst)}$~\cite{whelitevcombi}, in good
  agreement with the SM prediction.

\subsection{Top Quark Mass Combination}
\begin{figure*}[t]
\centering
\includegraphics[width=50mm]{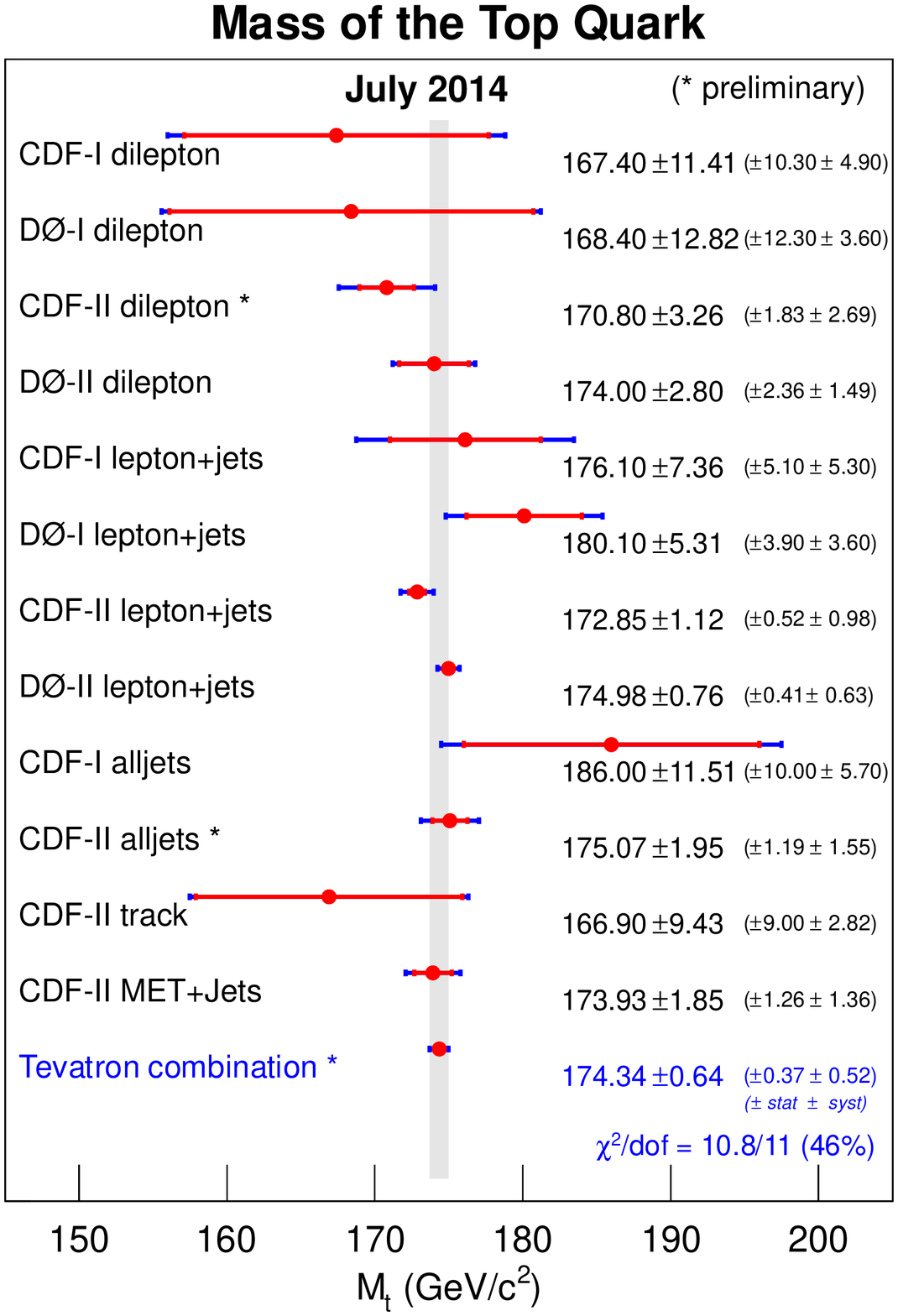}
\includegraphics[width=95mm]{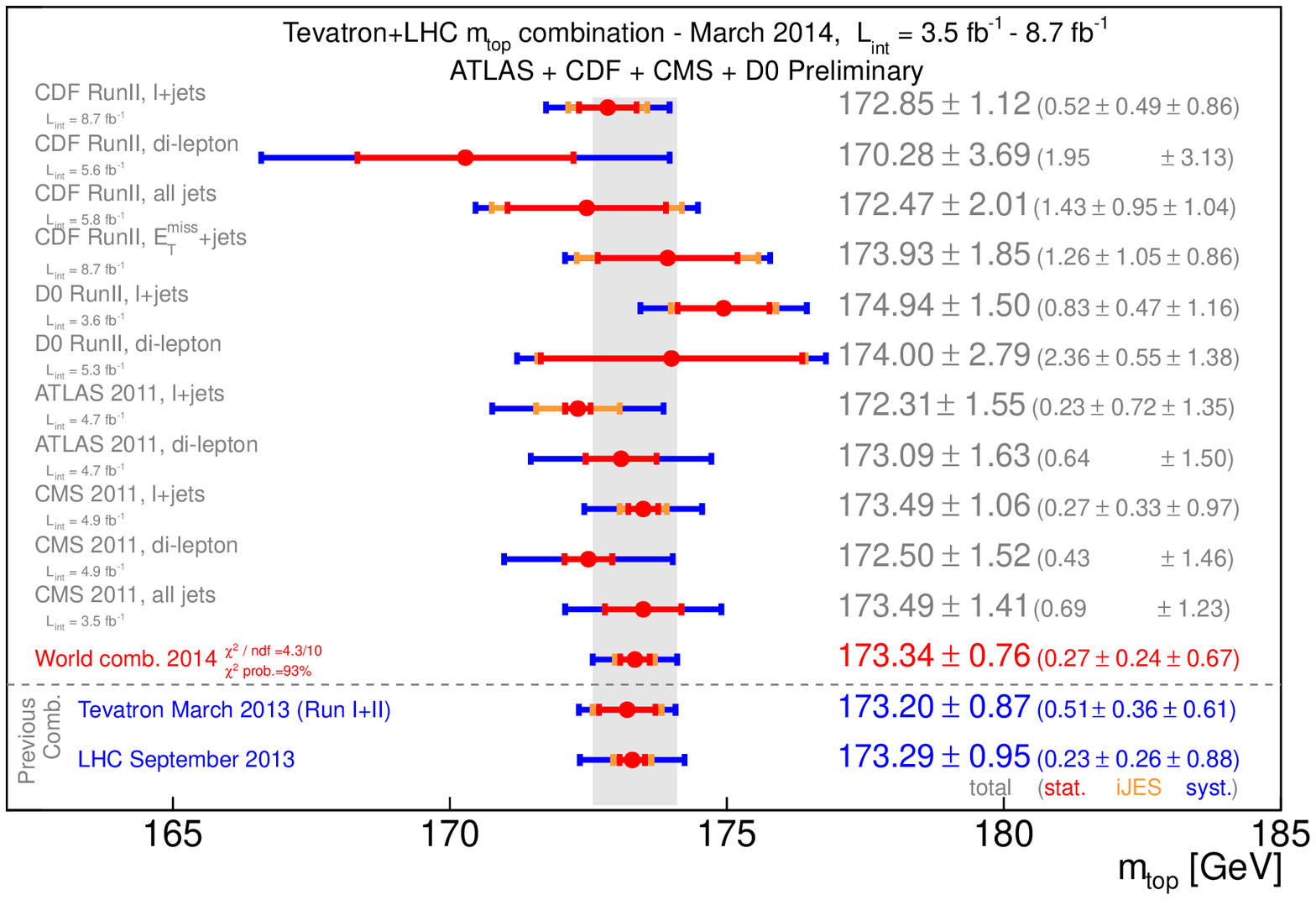}
\caption{ Recent Tevatron combination of the top-quark mass
  (left)~\cite{tevnewcombi} and world top-quark mass combination (right)~\cite{worldmasscombi}.} \label{fig:masscombi}
\end{figure*}

The mass of the top quark is an important free parameter of
the SM.  In order to get the most precise value,
combinations of the different analyses within each experiment, and
 between the experiments are necessary. The main challenge for the
combination is the proper handling of the correlations between the
systematic uncertainties, given that the total uncertainty on the top-quark mass measurements is often dominated by the systematic
uncertainty. 
For the latest Tevatron combination, five measurements from Tevatron
Run~I are used, as well as five published analyses from Run~II and two
preliminary results by the CDF collaboration. A first published
Tevatron combination, where only published results were used as inputs,
was released in 2012~\cite{tevatroncombi}, using up to
5.8~fb$^{-1}$. Since then CDF updated their analyses in the dileptonic
and fully hadronic final states, while D0 performed an updated analysis
in the semileptonic final state, using the full Run~II data
sample. The procedure follows that of the published
combination: The tool used for the
combination is BLUE; the systematic uncertainties are categorised in
classes according to a common source and correlations. The new
Tevatron top-quark mass combination, using up to 9.7~fb$^{-1}$ of
data, yields $m_{top}= 174.34 \pm 0.37 {\rm (stat)} \pm 0.52 {\rm
  (syst)}$~GeV~\cite{tevnewcombi}. The main contributions to the
systematic uncertainty arise from signal modelling and the uncertainty
on the in-situ light-jet calibration. Figure~\ref{fig:masscombi}
(left) shows the Tevatron top-quark mass combination and all the
individual input measurements.

\section{World Top Quark Mass Combination}
Once that LHC~\cite{lhccombis} and Tevatron had performed separate combinations of the top-quark mass measurements, the next natural step was to perform the
first world combination of the top-quark mass, which happened in March 2014. In order to ease the
treatment of correlations, the inputs to the combination are the
respective  best measurement per channel per experiment. Based on this
principle, the
combination uses six measurements from the Tevatron and five from LHC
as input. In particular, from CDF a measurement using a template
method in the semileptonic and a measurement based on neutrino
weighting in the dileptonic channels are used. From D0, a measurement
using the Matrix Element method in semileptonic and also a neutrino
weighting technique in the dileptonic final state are used, where the
in-situ jet energy scale calibration from the measurement in the
semileptonic channel is transferred to the measurement in dileptonic
events. The inputs contributed by ATLAS are a 3D template method in
semileptonic events and a measurement based on an $m_{\ell b}$
observable in dileptonic events, while from CMS an analysis in the
semileptonic and the fully hadronic final state, using the ideogram
method, are used as input, as well as an analysis in dileptonic events
based on analytic matrix weighting. All analyses by the ATLAS and CMS
collaboration are based on the $\sqrt{s}=7$~TeV data.

In the same light as for the Tevatron mass combination, the main challenge for the
world combination  is the treatment of systematic
uncertainties. Given that the four experiments have different methods to estimate their systematic uncertainty,
the situation is even more complicated and a proper classification and
estimation of the correlations is crucial. 
The current solution to this problem is to classify the
systematics into logical classes with the same correlation, determine
the central result for this choice, and then vary the correlations to
check the stability of the result. Several of these checks
have been done, where either a global scale is applied on all
correlation coefficients of experiment-only, LHC-only, Tevatron-only,
collider-only or all-experiments correlation factors, or where for
individual classes the correlation values are varied, and the impact
on $m_{top}$ or the uncertainty on $m_{top}$ are checked. It turns out
that the size of the effect of these different choices is quite small, and thus
no additional uncertainty is assigned for the choice of correlations. 
The combined world top-quark mass is $m_{top}= 173.34  \pm 0.27 {\rm (stat)} \pm  0.71 {\rm (syst)}$~GeV~\cite{worldmasscombi}. In Figure~\ref{fig:masscombi}
(right) the world mass combination and the inputs from all experiments
are listed. This result does not include
the latest results from Tevatron and LHC, which will reduce the
uncertainty even further, but also require even more care to be taken on the
treatment of systematic uncertainties and correlations. 

\section{Conclusion}
Several Tevatron top-quark combinations have been performed, as well
as the first world combination. Especially the observation of
$s$-channel single top-quark production, which was only possible by
combining the CDF and D0 analyses, shows the value of combining results
from different experiments. At both colliders, further combinations of top-quark
properties are planned, and will help to understand the
heaviest known elementary particle as precise as possible.

\ack
I would like to thank our collaborators from ATLAS, CDF, CMS and D0 for their help in preparing the
presentation. I also thank the staffs at Fermilab, CERN and
collaborating institutions, and acknowledge the support from ERC and the Helmholtz association.


\section*{References}
\begin{thebibliography}{9}
\bibitem{cdftopdiscovery} F.~Abe {\it et al.}  [CDF Collaboration],
   Phys.\ Rev.\ Lett.\   {\bf 74}, 2626 (1995).
\bibitem{d0topdiscovery}  S.~Abachi {\it et al.}  [D0 Collaboration],
   Phys.\ Rev.\ Lett.\   {\bf 74}, 2632 (1995).
\bibitem{atlas} ATLAS Collaboration,  JINST {\bf 3}, S08003 (2008).
\bibitem{cms} CMS Collaboration,  JINST {\bf 3}, S08004 (2008).
\bibitem{tevschannel} T. Aaltonen {\it et al.} [CDF and D0
  Collaborations],    Phys. Rev. Lett. {\bf 112} , 231803 (2014).
\bibitem{tevsinglecombis} T. Aaltonen {\it et al.} [CDF and D0
  Collaborations],    CDF note 11113, D0 note 6448 (2014).
\bibitem{blue} L.~Lyons, D.~Gibaut and P.~Clifford, Nucl.\ Instrum.\
    Methods in Phys.\ Res.\ Sect.\ A {\bf 270}, 110 (1988); L.~Lyons,
    A.~Martin and D.~Saxon, Phys. Rev. D {\bf 41}, 3 (1990); A.~Valassi, Nucl.\ Instrum.\ Methods in Phys.\ Res.\
    Sect.\ A {\bf 500}, 391 (2003).
\bibitem{tevxseccombi} T. Aaltonen {\it et al.} [CDF and D0
  Collaborations],    Phys. Rev. D {\bf 89}, 072001 (2014).
\bibitem{nnloheli} A.~Czarnecki, J.~G.~Korner, J.~H.~Piclum,
  Phys.\ Rev.\  D {\bf 81}, 111503 (2010).
\bibitem{whelitevcombi} T. Aaltonen {\it et al.} [CDF and D0
  Collaborations],   Phys. Rev. D {\bf 85}, 071106 (2012).
\bibitem{tevatroncombi} T. Aaltonen {\it et al.} [CDF and D0
  Collaborations],   Phys. Rev. D {\bf 86}, 092003 (2012).
\bibitem{tevnewcombi} The CDF and D0 collaborations, arXiv:1407.2682
  (2014). 
\bibitem{lhccombis} For LHC combinations see Giorgio Cortiana's
  contribution in these proceedings.
\bibitem{worldmasscombi} The ATLAS, CDF, CMS and D0 collaborations,
  arXiv:1403.4427 (2014). 

\end{thebibliography}
\end{document}